\documentstyle[ twocolumn, aps, epsf]{revtex}

\begin{document}

\draft \title{How to teach statistical thermal physics in the introductory physics course}

\author{ Koo-Chul Lee }

\address{Department of Physics and Center for Theoretical Physics,      Seoul National University,               
 Seoul, 151-742, Korea} \date{\today}
\maketitle 

\widetext 
\begin{abstract}

We report several simulation programs$^1$ which can be used to teach the statistical foundation of thermal physics in the  introductory college physics courses.  These programs are simple applications of a technique of generating random configurations of many dice with fixed total value.  By simulating dice throwing only  we can demonstrate all the important principles of classical thermodynamics.   

\end{abstract}

\narrowtext 

\section{Introduction}

The laws of thermodynamics were 
formulated long before we knew the existence of 
atoms and molecules.

The formal thermodynamics curriculum 
based on the empirical laws have been replaced 
by courses based on physical  and statistical 
 foundation of the subject in the intermediate 
courses in undergraduate curriculum for some 
time now.  In these courses they show how the 
properties of macroscopic system are simple 
consequences of the behaviors of their 
elementary constituents. Unfortunately, however, 
in the introductory physics courses the 
``traditional" exposition of the subject following the 
historical sequence of events is still prevailing. 

In the age when a single atom 
can be trapped and manipulated teaching thermal 
physics starting from the empirical laws of 
thermodynamics is a pedagogical scandal.   It is a 
high time to reform the style of teaching the 
thermal physics in the introductory college 
physics course.  

In 
most of introductory college physics text books 
authors introduce the laws of thermodynamics 
before going on some brief exposition on the 
statistical basis of subject if they mention it at all.
They introduce the concept of entropy 
nearly at the end of thermal physics part using 
Clausius'  definition 
\begin{equation}
 dS = dQ/T   \label{eq:clausius}
\end{equation}

\noindent
in conjuction with   the  Carnot cycle.  On the other hand 
Boltzmann's definition of the entropy 

 \begin{equation}
S = k\ln W \label{eq:boltzmann}
\end{equation}
\noindent
is usually appear as if a passing remark with brief 
comments.  Almost none makes any attempts to 
relate two expressions.  
The Clausius' definition of entropy is one of the  most difficult 
subject to teach.   First of all  $T$  in 
the above expression (\ref{eq:clausius})  is the absolute temperature. 
 If it were not the absolute temperature  in the expression (\ref{eq:clausius}) the Clausius' definition is meaningless. 
However  the concept of the absolute temperature has 
never been fully explained previously in 
textbooks that follow traditional style.   
The temperature is usually  introduced through 
operational  definition which can not explain the 
significance
\\*
\\[2.0cm]  of the absolute temperature other than  a curious
 experimental fact.  We can not convey the significance
 of the absolute temperature without any inkling of statistical nature of the absolute temperature.

Secondly  $dQ$  is usually introduced in conjuction with the 
first law of thermodynamics where it is implied an 
arbitrary infinitesimal quantity of heat.  However in the 
Clausius' definition  (\ref{eq:clausius})  the left side is  an 
exact or total differential of a state function $S$.   How can we  expect students understand meaning of the Clausius'  expression if we do not explain them carefully and clearly the  physical and 
mathematical circumstances which let us connect  an arbitrary infinitesimal quantity to an exact differential of a some function, $S$.   This subject is not an easy topic even in the intermediate thermal physics courses, where we devote much more time for the discussion of the subject. 

In many  cases instructors bypass all these details of the 
Clausius' expression and proceed to apply the 
formula to thermodynamic processes.  

On the other hand Boltzmann's definition is simple and clear enough.  Only problem is how to relate the Boltzmann's expression to  thermodynamic functions in a simple and clear manner. It is the task of statistical mechanics but statistical mechanics  is generally considered too difficult subject to teach in the introductory physics courses. 

In this paper I propose that  teaching statistical mechanics in the introductory college physics courses  is not at all difficult   if we employ a proper method. The proper method I propose in this paper is a simulation of dice throwing  on computers. Dice are familiar objects in our daily life and  any students have an experience of throwing dice.  Students can learn how thermodynamic principles emerge if they throw a large number of dice,  say tens of thousands of dice.  Computers came of age to let students throw many dice.   With the aid of computer simulations students can easily understand the statistical foundation of thermal physics.

 A Chinese 
proverb says$^2$: 

``  I hear, I forget;~~
  I see, I remember;~~
  I do, I understand." 
 
There is no better way than letting students run the simulations themselves and understand them. I wrote all the programs discussed  in this paper in JAVA applet and can be viewed and run at the web site referenced in the Ref.1 with a JAVA enabled browser such as  Netscape (version 4.04 with patches  or later version)  or MS Internet Explorer (4.0 or later version).

Students should be aware of the fact  that  the numbers we deal in statistical mechanics are  huge and are not even comparable with astronomical numbers as we demonstrate  in the conclusion of this paper.  This fact is intimately connected to a core principle of the thermodynamics, namely the irreversibility. This fact  is also important mathematically.   For example  if we throw $N$ dice simultaneously the number of possible outcome is  $W=N^6$.  Students who are accustomed to the linear quantity, a property of an exponential quantity, $W$ for a large $N$ such as
 \begin{equation}
   W \cong NW \cong \frac{W}{N} \label{eq:approximation}
\end{equation}
 may look like  paradoxical. However the validity of an  approximate relationship  (\ref{eq:approximation}) within the error of $O(1/N)$ for a quantity $\ln W/N$  is easily 
tested for the thermodynamic system where a typical $N$ 
 is of order of $10^{23}$.  However for small 
numbers such as $N$  = 3 or 4  the approximation 
(\ref{eq:approximation})  is certainly a nonsense.
The  validity of statistical mechanical argument begins at the point where the size of our system becomes large enough to tolerate the approximate relationship  (\ref{eq:approximation}).  
Students  can learn this fact  through  the computer simulations.

The dice system is a good analogy of a 
physical system made of $N$ paramagnetic atoms 
of spin $J=5/2$  in the external magnetic field$^3$.
 Atoms assume $2J+1 = 6$ equally spaced energy 
states.  We can make the spacing exactly unity by 
controlling the external magnetic field. Students need not know the detail of paramagnetic atom. It suffices to say that dice represent atoms that assume only 6 equally spaced energy levels.  Therefore the total value of dice is the total energy of the system. In all our simulation programs except the first one we label the value of dice from $0$.

In this simulation programs I emphasize 
that there are only two principles involved; the 
conservation of energy which is inherited from 
mechanics and the "principle of equal a priori 
probability" of accessible microscopic states 
consistent with the macroscopic specification. 
 We will put the conservation of energy by hand. This means we are considering the microcanonical ensemble.
 
 Therefore the first law is not a new thing but a 
simple application of the conservation of energy 
to the thermal system. 
In order to implement the ``principle of 
equal a priori probability" we use the 
microcanonical Monte Carlo (MC) technique first introduced  by M. Creutz$^4$  and further elaborated by myself$^5$.  
 
 The 
microcanonical MC technique is a simple 
device which allows us to scan uniformly all accessible 
microscopic states  of fixed total value or total energy.

We generalize common die to a die of $\nu$    faces 
where $\nu$  can be any integer numbers, $2 , 3, \cdots ,6, \cdots $.
 Although dice of arbitrary faces do not exist in the real world, one can imagine dice in the shape of tetrahedron($\nu = 4$) octahedron($\nu = 8$) etc.
Since dice of polyhedron shape are analogous to atoms of arbitrary number of equally spaced energy levels.

Other parameters of the system we can control in the 
simulation programs are the number of dice 
$N$  and  the  total value of dice $T$ which is the total energy $E$ for the corresponding  analogous physical system. 

The number $N$  is 
restricted to a square of an integer $L$ for programming convenience.  $L$ ranges 
from $2$ to any number restricted only by the 
computer resources. However for $L$ value  larger than $200$ ( $N>40000$) 
the dice configurations will  not be shown.  

Noninteracting atoms can not relax to thermal 
equilibrium state like an ideal gas.  We need to 
introduce a weak interaction.  The microcanonical 
MC technique used in the simulation 
program  introduces a mechanism which 
effectively amounts to physical interaction 
between magnetic atoms.  Suppose the spins of the magnetic atom
couple to the phonons of the underlying lattice so 
weak that only a single phonon interact with one 
magnetic atom at a time.   In the program a single 
atom is randomly selected and exchange energy 
with the phonon either by taking in or giving off energy 
according to a specified rule. 

In the next section we will explain the microcanonical MC technique by way of
simulating rolling of dice.   
In order to make students feel at home we 
throw only dice  in the simulation programs and a phonon is represented by the Maxwell's demon$^4$ who carries a bag of energy. 

Following three sections we present 
dynamical entropy simulation, statistical 
temperature simulation, Maxwell-Boltzmann 
distribution simulation. In all these simulations we 
throw only dice and watch the consequences. In 
the final section we make a summary and 
concluding remarks. 

\section{Principle of equal a priori probability and  microcanonical Monte Carlo technique}
\label{sec:two}

           The fundamental postulate of equilibrium statistical mechanics is the
principle of equal a priori probabilities. 

          To college freshmen it suffice to say that the internal molecular
interaction is complex enough that all microscopic states consistent with the
macroscopic specification are equally likely to occur. This is analogous to the simple
postulate of assigning equal probabilities to head and tail in an experiment of
tossing a coin.  The ultimate validity of the postulate can only be established by
checking its predictions against experimental observations.

          In order to realize the postulate we adopt the microcanonical Monte
Carlo technique$^4$  which has been successfully used for calculating thermodynamic
functions of more complex system with precision and efficiency$^5$. 

          We can do this by simulating a throw of $N$  dice on computers.  The
total value designated by $T$ of dice which is the sum of the value of individual dice.  We throw dice for a fixed $T$. 
Since $T$  is   the total energy of the analogous physical system we are generating a microcanonical ensemble. 

          How can we throw $N$  dice whose total value matching a some
predetermined value $T$?  In other words how can we generate a random dice
configuration of a given $T$?   We can think of throwing dice randomly until we get a
throw whose total value matches $T$.   This is impractical unless either the number of
dice is very small or $T$  is close $3.5N.$ (If we throw  $N$ dice randomly we will most
likely get $T$   values which is close to $3.5 N$.)  The microcanonical MC
technique is one efficient method of generating random configurations of dice for a
given fixed  value of $T$. 

For the purpose of illustration let us consider a system of $N = 100$ dice and generate a microcanonical ensemble of $T = 260$.  

          Our  system consists of  $100$  dice and a demon who
carries a bag of maximum capacity of $10$ units of value or energy. 
There are $5$ units
of energy in the demon's bag initially as in the Fig.\ref{fig:demon1}. Therefore the total value of dice+demon
system is $265$ and this number remains fixed throughout the simulation.   In
other words the system made of dice and the demon is a closed system. 
\begin{figure}
\epsfxsize=7.97cm
\epsfysize=5.93cm
\hskip-0.1cm
\vskip0.1cm
\Large
\centerline{Fig.1}
\normalsize
\vskip0.7cm
\caption{The initial configuration of the    applet 1. if   you ``Select \& Roll" button you will   get a randomly selected dice on the
  margin to the right side and the new  face as a result of the roll. The amount of  increment or decrement in value as a result of the roll  is indicated. Examining the state of bag, the  decision whether accept the new configuration or not  is also indicated by "Yes" or "No".  By clicking "Replace or Put it back" button you get a new configuration or retain the old one.}
\label{fig:demon1}  
\end{figure}

          Thermalization or MC step consists of following steps: (1)
The demon selects one die out of 100 dice randomly and rolls it. (2) The new
configuration generated by the rolled die makes the total value of dice system
either increase or decrease. If the increment can be covered by the energy units in
the bag or decrement can be accommodated by enough vacancy in the bag the new
configuration is accepted. Otherwise the move is rejected and the die retains the old
value and no new configuration is generated. (3) The demon repeats the same
procedure, steps (1) and (2).

          This MC procedure amounts to the random walk in the
configurational  space(analogous to the phase space for the Hamiltonian system)
bounded by the energy band  $254 < T <266$ which is analogous to the energy shell
for the continuous Hamiltonian system.  We define one thermalizing MC
step by whole procedure until the new configuration of the given $T (=260)$ is
generated.  This procedure satisfies the condition of detailed balance and can be
made ergodic$^4$.

          It should be noted that the capacity of the bag is the minimum size to
keep the successive two configurations of the given $T$ from being the same one.  At $T=260$ the value in the bag is $5$ units and there are $5$ units of vacancy. Therefore any change in the value of a single die can be accepted always. Some times the rolled dice turn up the old value again but it is still considered as a fresh configuration although the configuration is unchanged.  On the other hand if the
bag size is larger than $10$ , it will take much longer time to return to the given $T=260$
state while for a bag of capacity  smaller than $10$  there is a chance that the
successive two configurations might be the same one by rejecting the move in the step
(2) making the statistics deteriorated$^5$.
\begin{figure}
\epsfxsize=8.04cm
\epsfysize=5.96cm
\hskip-0.1cm
\vskip0.1cm
\Large
\centerline{Fig.2}
\normalsize
\vskip0.7cm
\caption{A random dice configuration with  $T=260$ which is obtained from the initial configuration of Fig.\ref{fig:demon1} after performing  several MC steps. 
}
\label{fig:demon2}  
\end{figure}

Fig.\ref{fig:demon2}  
 is a typical configuration obtained after a few MC steps. 
    In the following and subsequent chapters all the simulations use this
thermalizing MC steps and the demon will be hidden in the background and
configurations of $T \neq 260$ states will also be suppressed and hidden.  

\section{Thermal relaxation and change of entropy }
 \label{sec:three}

           Starting from the initial configuration given by the dice arrangement
shown in Fig.\ref{fig:demon1} of the last section we thermalize the system by performing the
MC steps and watch how the configuration changes in time. The time is the
number of MC steps described in the last section. 

           Now we calculate the entropy by the Boltzmann's celebrated
formula given by the eq. (\ref{eq:boltzmann}).  $W$ in the eq. (\ref{eq:boltzmann})
 is the number of configurations and the $k$ is the Boltzmann's constant
which defines  the unit of thermodynamic entropy.  In this paper we set $k = 1$. 

          From this point onwards we relabel the face of dice as $\{0, 1, 2, 3, 4, 5\}$  and use different colors to represent these values  instead of ``dots".  This
will make the total value $T$ of the last section change to $T-N$. 

          If we take a view that the  total value $T$  is the only macroscopic
parameter that specifies our macroscopic state then $W = W(T)$, where $W(T)$  is the total number of
configurations that have the total value $T$.  Then the {\em equilibrium} entropy of the
system is 
\begin{equation}
S(T) = \ln W(T). 
\end{equation}

           Let us subdivide the microscopic configuration into classes of configurations
having  a  fixed  set  of  values   
$\{ n_0,  n_1,  n_2,  n_3,  n_4,  n_5\}$ 
which are the number of dice
showing  $0, 1, 2, 3, 4,$ and $5$ respectively.  We will call these numbers ``occupation
numbers''.  We designate the number of configurations belonging to a fixed set of
occupation numbers, $\{ n_0,  n_1,  n_2,  n_3,  n_4,  n_5\}$   by $W(\{n_k\})$.  Then 
\begin{equation}
W(T) = \sum_{\{n_k\}}  {^\prime} W(\{n_k\})    \label{eq:entropy}
\end{equation}
where `` $^\prime$ '' in the above summation implies that the sum is carried out over all
possible set of values $\{ n_0,  n_1,  n_2,  n_3,  n_4,  n_5\}$   consistent  with the condition
\begin{equation}
 \sum_{k=0}^5  n_k = N    \label{eq:normalization}
\end{equation}
and 
\begin{equation}
 \sum_{k=0}^5  kn_k = T   \label{eq:average}
\end{equation}

In the expression (\ref{eq:entropy}) the number of summand is  $O(N)$ (at most $N^6$ in this case).   This implies that  at least one of the summands must be $O(\exp(N))$  if $W(T)$ is  $O(\exp(N))$  in view of the approximate relations (\ref{eq:approximation}).  This then allow us to replace  the sum (\ref{eq:entropy}) by a single largest term.
{\it i.e.}
 \begin{equation}
W(T) = W(\{\tilde n_k\}).    \label{eq:eq-ent}
\end{equation}
        
$W(\{\tilde n_k\}) $ in the above expression(\ref{eq:eq-ent})  is the largest term among the summands in the  eq.(\ref{eq:entropy}).

Let us now generalize the equilibrium entropy into a generalized entropy which represents a nonequilibrium state by
\begin{equation}
 S^{gen} = \ln W(\{n_k\}) ~.  \label{eq:neq-ent}
\end{equation}

          Since the $W({n_k})$  of the generalized entropy (\ref{eq:neq-ent})  is nothing but the multinomial coefficient given by 
 \begin{equation}
 W(\{n_k\}) = \frac{N!}{n_0!n_1!\cdots n_5!}~ ,  \label{eq:multinomial}
\end{equation}
   we can write  the generalized entropy  (\ref{eq:neq-ent}) as 
\begin{equation}
 S^{gen} = \ln N! - \sum_{k=0}^5\ln n_k! ~.  \label{eq:multi-ent}
\end{equation}

In the simulation programs the entropy is calculated using the formula (\ref {eq:multi-ent}).

Fig.2a is the default initial configuration of the applet 2 which  is the same dice arrangement as in the Fig.\ref{fig:applet2i}  
 of  the last
section  and  Fig.\ref{fig:applet2e} is the typical simulation result of the applet 2 with a different set of system parameters. 
\begin{figure}
\epsfxsize=8.22cm
\epsfysize=6.17cm
\hskip-0.1cm
\vskip0.1cm
\Large
\centerline{Fig.3}
\normalsize

\vskip0.7cm
\caption{The default initial  configuration of the applet 2 which is exactly the same configuration as shown in the Fig.\ref{fig:demon1}.  If we click the   ``Start/Resume" button we generate  a series of configurations  such as shown in the Fig.\ref{fig:demon2}. 
}
\label{fig:applet2i}  
\end{figure}

       We see  in the Fig.\ref{fig:applet2e}, the entropy defined by eq, (\ref {eq:multi-ent}) approaches to the equilibrium entropy as time progresses. 
 Students can watch how the configuration changes visually and the computer calculate the entropy at every instance and plot a graph.  
 
         As the time progresses we will see the ``entropy" increase
monotonically and hit the maximum value eventually and remains there ``forever"
apart small fluctuation which tends smaller as the system size grows larger.
 This is  the 2nd law of thermodynamics ! 

 \begin{figure}
\epsfxsize=7.94cm
\epsfysize=5.93cm
\hskip-0.1cm
\vskip0.1cm
\Large
\centerline{Fig.4}
\normalsize

\vskip0.7cm
\caption{The typical simulation result of the applet 2.  The scale of the energy density  is (0.0 - 5.0). Initially the upper region starts out with the energy density 5.0 while the lower region starts out with 0.0. 
}
\label{fig:applet2e}  
\end{figure}

  The applet 2  also   calculate   the      energy    density of the two   regions,     namely   the upper                    region    where the      dice of  value 6(now 5)  were   initially  occupied  and the  remaining  lower                                region   where   dice of   value 1(now 0)  were  occupied.     We will    see that   the MC              steps   randomize    the  configurations in both  ``energy"   and   ``configurational"   space. As                                                 the  system relaxes to  equilibrium   state the energy    density    becomes  uniform                             everywhere   manifesting  one characteristic  of  equilibrium state. 

          In the right lower corner in the applet 2, the equilibrium entropies are plotted. The graph shows the thermal fluctuation in the equilibrium state in an enlarged scale in changing colors.  The  zoomed-in scale is marked in the left side of the original plotting region.  Students can estimate the size of the fluctuation as the size of the system grows.

           On the other hand in the opposite limit where the number of dice is small, say 4, then average
period of returning to the initial state is less than $64 = 1296$  which students  can observe
during the experiment. By estimating these probabilities students can understand
the meaning of irreversibility.

            In the
applet 2  students can edit 3 system variables namely  the energy density, $e$ which is
 $T/N$,  the size of the system  by $L$, which is the square root of $N$,  and number of
energy levels, $\nu$  which is the number of faces of a die ($6$ for an ordinary die).  ``$e$"
accepts a real number and is converted to a closest integer $T$  by multiplying  $N$  with a
possibility of a loss of precision.

          There are two more editable variables in the applet 2: They are ``time",  the number of thermalizing MC steps and the ``NoDP" , which is the number of data points
to display.   Therefore every  ``time/ NoDP"  MC steps generated data are displayed or  plotted. 

          For a given $T$,  the initial state is constructed such a way that $T$ is made of $q$ dice of highest value $\nu-1$ and single die of value $r$ , where $q $ is the quotient and $r$ is the remainder satisfying 
$T=(\nu-1)q+r$.   The priority is given such a way that dice of higher value occupy the top and leftmost  positions. 

          Finally the applet also plot the entropy calculated using approximate formula
\begin{equation}
 S^{Stir} = -\sum_{k=0}^{\nu-1}\frac{n_k}{N}\ln \frac{n_k}{N},   \label{eq:stirling}
\end{equation}
in green color.

$S^{Stir}$ is calculated using the Stirling's approximation for the factorial,

\centerline{ $\ln n! = n\ln n - n$ } 

\noindent 
which is made in the same spirit as the approximation (\ref{eq:approximation}). We plot this to show the range of the system size where the statistical argument based on this approximation becomes valid.  

           The points to watch in this simulation are:

              (A) The visual characteristics of thermal equilibrium state such as
    uniformity in various densities. 

              (B) The relaxation, irreversibility, 2nd law of thermodynamics.
    Although the algorithm described in the above is periodic (any simulation on the computer is periodic for that matter),  if N is moderately large number, say
    N=100, there is absolutely no chance that we see the configuration return
    to the original unique arrangement or close to it.  This is the irreversibility. It
    is sheer enormous length of the recurrence period (Poincar\`{e} cycle) which
    makes the reversibility impossible. 

              (C) The size dependence of fluctuations.  Students can watch the fluctuation
    subsides as the system size grows keeping all other parameters intact.
    Students can watch  the fluctuations in  
    the entropy  and local energy densities shrink proportionally to 
      $1/\sqrt{N}$  as $N$ grows. 

              (D)The entropy versus the energy. The entropy as a function
  of  energy will be shown in the next section. However  in this simulation  students may try various energy densities ``$e$" to see how the equilibrium entropy changes as a function of energy . Since the 
    energy levels of this system is finite the equilibrium entropy does not always increase as the energy density increases.  Some students may discover the concept of negative absolute temperature. 

              (E)The extensibility of the entropy.  Students can watch how the entropy
    density change or remain as the size of the system grows. 

\section{Thermal equilibrium and statistical temperature }
\label{sec:four}
 
           In the last section we presented a simulation program, the applet 2 which demonstrates the process of relaxation of a single system  to the equilibrium state and the characteristics of the equilibrium state. In this section  we present a simulation program that demonstrates the  behavior of
relaxation of two systems which are  thermally interacting with each other. From this point  onwards we
denote the total value of dice by $E$ instead of $T$ and $T$ will be used to denote the
absolute temperature.

          Initially we separate the two systems {\bf A} and {\bf B} by a thermally
insulating wall so that they can not exchange energy. We prepare our system in highly
nonequilibrium states for both systems as described in the last section. When we thermalize  two systems with
insulating wall on, then the relaxation processes are just the same as in the case of
a single system.  When we remove the wall two systems can exchange energy.
 Simulation program is designed such a way that two independent microcanonical
MC processes of fixed energies are performed,  when two systems are insulated,  alternating one MC step each for {\bf A} and {\bf B}.  When the wall is removed
we use the single microcanonical MC process.  In this case we also perform MC
steps alternately for {\bf A} and {\bf B}  parts.
\begin{figure}
\epsfxsize=17.22cm
\epsfysize=13.69cm
\hskip-0.1cm
\vskip0.1cm
\Large 
\centerline{Fig.5}
\normalsize
\caption{The typical simulation result of the applet 3.  The entropy and energy densities  as well as the equilibrium canonical inverse temperatures for {\bf A} and {\bf B} parts are plotted in different colors in the same rectangle in upper right corner. The scales of these quantities are shown in the middle center area.  The scales of the plotting area for the entropy vs. energy is the same one as the scales of the main plotting area.   At every plotting instance the values of these data are also displayed during the simulation in the upper middle area.  The total entropy and energy densities are simply $(S_A+S_B)/(N_A+N_B)$ and  $(E_A+E_B)/(N_A+N_B)$.
}
\label{fig:applet3}  
\end{figure}

          In the applet 3(Fig.\ref{fig:applet3}), we can  control  the size,  $N_A, N_B$ the
number of energy levels $\nu_A, \nu_B$, initial energies, $E_A, E_B$ for two systems {\bf A} and {\bf B}  separately.   There
are only one length of time of simulation and a single number of data points. 

    During  the  simulation  process   we  can  toggle    the  insulating  wall  on  and  off  by                           clicking  the check box.      This  will    make   the system   relax  to  the final equilibrium             state  in   may    different    paths.

          The equilibrium condition for two system in thermal contact where the
energy exchange between two systems are allowed is given by the maximum
entropy principle.  For  if we maximize  total entropy 

\centerline{ $ S = S_A(E_A) +S_B(E_B) $ }  \noindent
    under the condition that the total energy of combined system
 \centerline{ $ E = E_A +E_B  = constant$, }  \noindent
we obtain 
\begin{equation} 
     \frac{dS_A}{dE_A} =\frac{dS_B}{dE_B}~.  \label{eq:sbye}
\end{equation} 

          We should emphasize the fact that the equilibrium condition for two systems
in thermal contact, is that two slopes of entropy as a function of energy equal to each other.  This simulation program explicitly  shows  that the equilibrium is
reached when the slopes of two entropy functions $S_A$  and $S_B$  become equal. 

          Therefore statistical definition of the inverse temperature  $1/T$ naturally follows as
\begin{equation} 
     \frac{1}{T} =\frac{dS}{dE}~.  \label{eq:invtemp}
\end{equation} 

Of course any single valued function of the entropy slope may be defined as the temperature.  However the absolute temperature was known before the statistical mechanics was discovered. In order to make the previously known absolute temperature such as the one that enters in the  Clausius entropy expression (\ref{eq:clausius}) coincide with the statistical definition,  the temperature must be defined as in the eq.(\ref{eq:invtemp}) as we all know.

          This temperature can be
calculated during microcanonical MC simulation by examining how often the system visit the energy
levels  $E-1$, and $E+1$. If in a given period of simulation the frequency of visits are
recorded as $f_-$  and  $f_+$,  then  $S_+ - S_-  = \ln(f_+/ f_-)$  and $T$ can be estimated by 
\begin{equation} 
     \frac{1}{T} = (\ln\frac{f_+}{f_-})/\Delta E  \label{eq:micro-temp}
\end{equation} 
\noindent
where $\Delta E  = 2$  in our case. This is the microcanonical temperature.

          However this method of estimating the temperature is time
consuming since we have to collect about $1000$ data for $f_\pm$'s for a reasonable statistics
at every instance.  This makes the simulation almost crawl. Fortunately we have an analytic
expression for the temperature for large systems$^3$.  

          In the applet 3 the inverse temperature $1/T$($\equiv \beta$)  is the equilibrium
canonical inverse temperature $\beta$ calculated solving the following equation for $\beta$ for a given
set of variables $e (=E/N)$ and $\nu$,
\begin{equation} 
    e = \frac{\nu-1}{2} - \frac{\nu}{2}\cosh\frac{\nu}{2} \beta +\frac{1}{2}\cosh\frac{\beta}{2}~.  \label{eq:can-temp}
\end{equation}

          This is the canonical temperature for large systems.   It was tested and verified that the canonical temperature agrees within the statistical error bar with the equilibrium microcanonical temperature calculated using the formula (\ref{eq:micro-temp}),  if the size of the system is large enough so that the Stirling's approximation remains valid.

 The Fig.\ref{fig:temp}  is  a  typical  simulation result  of the applet 3.   There  are  many steps where  thermodynamic                                                                    functions remain flat.  These regions are   when   the  insulating   wall  is   on  so  that   each  system                                                                    is  in  thermal    equilibrium  state   separately.

          The entropy as a function of energy is traced and  plotted  in a small box in the lower center area.
 At the end of the simulation the straight line with the slope calculated by solving the equation (\ref{eq:can-temp}) is drawn extending from  the end points of two curves.   Students  can
visually  confirm that these slopes are  indeed the slopes of the simulated entropy functions at the final equilibrium state.

           Points to watch in this simulation program are:

          (A) Zeroth law of thermodynamics.

          Although we do not have 3 system in thermal contact we can do so  indirectly by placing  2 different systems in the position of {\bf B} in thermal
contact with one common system, {\bf A}.  

          (B) Students should watch the small box where the entropy as a function of energy is
being traced during the simulation. It shows the functions are  concave downward. If these two systems start at the points where their slopes are  different then they reach thermal equilibrium states
when their slopes become equal. If they overshoot  the
equilibrium point then it will make the total entropy decrease. It is a simple law of probability which keeps this from happening!     

           (C) Students can also make a rough estimate of  the entropy change in the process of heat
transfer, by calculating  $dS_A = dE_A/T_A$  and $dS_B = dE_B/T_B$    together
with the net change $dS = dE/(1/T_B-1/T_A)$, where $dE = dE_B =-dE_A$.  

 This can be
done by suspending the simulation just before removing the wall and after reinsert
the wall.  In this way students  can verify that the Boltzmann's entropy is indeed the same
quantity defined by Clausius(\ref{eq:clausius}),  $dS = dQ/T$, where $dQ = dE$. 

           (D)Students can experiment various paths to reach to the thermal equilibrium
state by drawing the curves over the old one to confirm that  the final equilibrium state is indeed a single
unique state.

\section{Maxwell-Boltzmann distribution }
                           
In most text books for  introductory college physics courses, authors introduce Maxwell-Boltzmann (M-B) distribution  and  discuss its consequences. However none derives or even present any plausible argument for the origin of the distribution.

If we throw and generate a truly {\em  random configuration} of sufficiently large number of dice  keeping the total value $E$ fixed and examine the  distribution of occupation numbers $\{\tilde n_k\}$, we  get the M-B distribution given by 
\begin{equation}
    \tilde n_r = A\exp(-\frac{r}{T}),  ~~r = 0,1,2, \cdots 5.  \label{eq:mb}
\end{equation}

Why?  It is not so difficult to explain the reason  to students in the introductory physics courses if students understood the probability, entropy, and temperature which we  discussed and demonstrated in the previous sections.   

Truly random configuration can be realized  when the system is in the thermal equilibrium.  Let us concentrate one die say, one in the upper leftmost corner.   Let us keep performing  MC steps after the dice system reached the thermal equilibrium and ask how often this particular die will show a value $r$.  This frequency $f_r$ must be proportional to the number of configurations of the rest of dice to have total value $E-r$,~ 
{\it i.e.}, $W_{N-1}(E-r)$. If $N$ is sufficiently large, $W_{N-1}(E-r)$ can be approximated by $W_N(E-r) = W(E-r)$.
Therefore 
\begin{eqnarray}
f_r \propto W(E-r) & =& \exp(S(E-r))\nonumber \\ \nonumber &\cong&  exp(S(E)-\frac{dS}{dE}r) \\  &=& A exp(-\frac{r}{T}). 
\end{eqnarray}
In the above derivation we have used  the definition of the temperature (\ref{eq:invtemp}), the Taylor series approximation $S(E-r) \cong S(E) -(\frac{dS}{dE})r$  since $r<<E$.

          This is the standard argument for deriving the canonical distribution$^6$   and  is not much more difficult than what we have done so far. 

In the applet 4, instead of examining the frequency $f_r$ of a single die which is time consuming we examine the distribution of occupation numbers $\{\tilde n_r\}$ since every die is equivalent and the outcome of other dice may be considered as the manifestation of the single die in question in the time series. Another words it is the demonstration 
\begin{figure}
\epsfxsize=17.04cm
\epsfysize=10.3cm
\hskip-0.1cm
\vskip0.1cm
\Large
\centerline{Fig.6}
\normalsize
\vskip0.7cm
\caption{The typical simulation result of the applet 4.  Applet 4 is very much the same as the applet 2.  We trace the change of occupation number 
$\{n_k\}$   in time instead of  the energy densities. At the end of the simulation when the equilibrium is reached we plot the bar diagram  of the $\{\tilde n_k\}$ using the colors representing dice values.
}
\label{fig:applet4}  
\end{figure}
\noindent of  the ergodic hypothesis that the time average is the ensemble
average! 

Fig. 4 is the typical simulation result of applet 4.  In this
  simulation   program,    applet 4   we   trace   the  distribution  of  occupation  numbers                                                             from   the  highly    nonequilibrium   state   to equilibrium  state.

          The curve drawn using cross mark ``x"   in the middle rectangle is
the entropy defined by the equation (\ref{eq:multi-ent}). It can be used  to determine if the system reached the equilibrium state. 

          The temperature used for drawing  the M-B distribution curve given by eq.(\ref{eq:mb})
is the canonical temperature defined in the equation (\ref{eq:can-temp}) and the constant $A$ is determined by the normalization condition (\ref{eq:normalization}). 

           Points to watch in this simulation program are:

          (A) Students can perform the experiments for systems of small number of dice and
increase the size gradually to see from what size the M-B distribution is applicable. 

          (B) Students can watch the fluctuation subsides as the size of the system grows.

(C) 
 The entropy of the equilibrium state is represented by  a single term
of a set of occupation numbers as we saw in the eq. (\ref{eq:eq-ent}). Therefore $\{\tilde n_k\}$ of the eq. (\ref{eq:eq-ent}) are the M-B distribution.  Students may discover  that the M-B distribution can also be derived  by maximizing $S^{gen}$ (\ref {eq:multi-ent})
 under the conditions (\ref {eq:normalization}) and (\ref {eq:average}).

\section{Conclusion}

          In this simulation programs I made the point that the statistical
foundation of thermal physics is not so difficult to teach to students in the
introductory physics courses if we have some simple examples to demonstrate or let
students try themselves. 

          The postulate of equal a priori probabilities is an easy concept  to
teach.  The energy conservation is not a new concept.  The crux is the realization of the postulate.   However if we use the microcanonical MC technique
we cam easily implement the postulate. Without the microcanonical MC technique it would  indeed be difficult to  generate a truly random configuration of fixed $E$.

The microcanonical MC is not a difficult technique to follow as we demonstrated in the sec.\ref{sec:two}.
Once students understand how to generate a truly random configuration of given fixed total value, $E$  then simulations that follow are simple applications of this microcanonical MC steps.

          Students can easily grasp the laws of thermodynamics especially the
2nd law together with the concept of irreversibility. The number of configurations
even for a  $100$ dice is $6^{100} > 10^{78}$ which far exceeds the age of universe in seconds
which is less than $10^{18}$.  Therefore it is impossible even for a system of $100$
dice in thermal equilibrium return to the initially prepared highly nonequilibrium
state.  Even if a computer generates a million configurations in a second the odd
that  we get the initially prepared nonequilibrium state in the age of universe is still
$10^{-54}$ ! 

          Students can also learn and understand that the entropy becomes
maximum at the thermal equilibrium and never decreases apart the fluctuations
which will get less as the system size grows. 

          From the maximum entropy principle of thermal equilibrium state it is
easy to understand the condition of two systems becoming  in thermal
equilibrium state.  It is when two slopes of the entropy as a function of
energy  become equal to each other.  From this understanding it is natural to introduce the statistical inverse 
temperature as the slope of the entropy as a function of energy. 

          The simulation also makes students understand the true meaning of
Clausius' definition of entropy. 
It is more natural to take a view that the Clausius' entropy formula(\ref{eq:clausius}) is defining the absolute temperature rather than defining the entropy. 

          In most introductory college text book M-B
distribution are introduced without derivation. Simply throwing many dice it is easy
to demonstrate and understand why occupation numbers must follow the M-B
distribution. 

          In conclusion it is not that difficult to teach the statistical foundation
of thermal physics in the  introductory college physics courses if we know how to teach
the statistical concepts. The microcanonical MC technique is one such
"how-to".

\acknowledgments

This work was supported in part by the Ministry of Education, Republic of 
Korea
through a grant to the Research Institute for Basic Sciences, Seoul National
University, in part by the Korea Science Foundation through Research Grant to
the Center for Theoretical Physics, Seoul National University.
 

\vspace{0.5cm}
\small
$^1$These programs were originally written to supplement a textbook(in Korean) used for the introductory college physics course  at Seoul National University, Seoul, Korea.  They are  translated into English  and uploaded at the web site: Koo-Chul Lee, ``Thermal Physics Simulations," 

\noindent http://phys.snu.ac.kr/StatPhys/frame-e.htm.

$^2$M. Johnson, D. Johnson, K. Lee, ``A Pedestrian Guide To Reforming The Introductory Physics Course,"  

\noindent http://www.phy.duke.edu/~mark/reform/.

$^3$F. Reif {\it Fundamentals of Statistical and Thermal Physics} (McGraw-Hill, New York, 1965),  pp. 257-262.

$^4$Michael Creutz, ``Microcanonical Monte Carlo Simulation,"  Phys. Rev. Lett. {\bf 50}, 1411-1414 (1983).

$^5$Koo-Chul Lee, ``A new efficient Monte Carlo technique,"   J.  Phys.  A:  Math.  Gen. {\bf 23}, 2087-2106  (1990).

$^6$See Ref.3   pp. 202-206.


\end{document}